\documentclass{article}

\usepackage{arxiv}
\usepackage{amsmath} 
\usepackage[utf8]{inputenc} 
\usepackage[T1]{fontenc}    
\usepackage{hyperref}       
\usepackage{url}            
\usepackage{booktabs}       
\usepackage{amsfonts}       
\usepackage{nicefrac}       
\usepackage{microtype}      
\usepackage{lipsum}		
\usepackage{graphicx}
\usepackage{natbib}
\usepackage{doi}

\title{A Multi-Embedding Convergence Network on Siamese Architecture for Fake Reviews}

\author{ \href{https://orcid.org/0000-0000-0000-0000}{\includegraphics[scale=0.06]{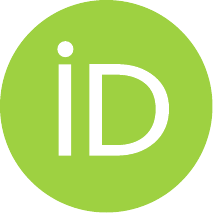}\hspace{1mm}Sankarshan Dasgupta}\\
	Department of Computer Science\\
	University of Dayton\\
	Dayton, OH 45469 \\
	\texttt{dasguptas2@udayton.edu} \\
	\And
	\href{https://orcid.org/0000-0000-0000-0000}{\includegraphics[scale=0.06]{orcid.pdf}\hspace{1mm}James Buckley} \\
	Department of Computer Science\\
	University of Dayton\\
	Dayton, OH 45469 \\
	\texttt{jbuckley1@udayton.edu} \\
}

\renewcommand{\shorttitle}{\textit{arXiv} Template}

\hypersetup{
pdftitle={A template for the arxiv style},
pdfsubject={q-bio.NC, q-bio.QM},
pdfauthor={David S.~Hippocampus, Elias D.~Striatum},
pdfkeywords={First keyword, Second keyword, More},
}

\begin{document}
\maketitle

\begin{abstract}
In this new digital era, accessibility to real-world events is moving towards web-based modules. This is mostly visible on e-commerce websites where there is limited availability of physical verification. With this unforeseen development, we depend on the verification in the virtual world to influence our decisions. One of the decision making process is deeply based on review reading. Reviews play an important part in this transactional process. And seeking a real review can be very tenuous work for the user. On the other hand, fake review heavily impacts these transaction records of a product. The article presents an implementation of a Siamese network for detecting fake reviews. The fake reviews dataset, consisting of 40K reviews, preprocessed with different techniques. The cleaned data is passed through embeddings generated by MiniLM BERT for contextual relationship and Word2Vec for semantic relationship to form vectors. Further, the embeddings are trained in a Siamese network with LSTM layers connected to fuzzy logic for decision-making. The results show that fake reviews can be detected with high accuracy on a siamese network for prediction and verification.
\end{abstract}


\section{Introduction}
Nowadays reviews have taken a major stride forward on online shopping decision and experience. An individual whether satisfied or not, most certainly shares an experience about the product on the website. A positive review has many benefits on the sale of the product, giving it an edge over other products to a new buyer and vice versa as per \cite{pang2008opinion}. This has also given rise to many competition in a particular domain in an e-commerce open market solution. This implementation to identify fake review demonstrates our commitment to social responsibility on manipulation damaging customer decision \cite{zhu2010impact}. Our motivation is to create a accessible design that ensures that technology is usable by the widest possible audience. In survey \cite{Survey_Fake_Reviews} many research approach are discussed to detect the unethical measures such as: fake reviews to influence the new buyer product selection. These algorithms discussed could eventually lead to positive user experience on e-commerce website for the customer. However, identifying the fake review is a conundrum, as these process could take a detailed understanding of the review patterns and their intentions. \cite{Knutsson:2022}  discusses on how some of the e-commerce websites have already taken steps to control the review process with tagging the review with a verified user. The manual tagging measures might give rise to many real review opinions being inconsiderate. In that scenario, failure to accept free information into consideration will lead to impact on new buyer's decision and damage the product's refinement.

Many new research has been conducted on identifying the reviews using deep neural networks \cite{alsubari2022data},\cite{bathla2022intelligent}. The algorithm's though has made advancement on this field but failed to demonstrate a universal acceptance on the algorithm. The reasons behind the failure might be the inaccurate detection due to diverse corpora of the text \cite{aladeen2023can}. Nowadays, artificially constructed reviews also consist of structured sentence and product description. These well studied fake review implementation needs a specific approach for identification. We propose an approach on learning algorithm to specifically target the fake reviews. In these proposed approach, we have used the fake review dataset \cite{salminen2022creating} readily available consisting of about 40K reviews. We clean the reviews data with techniques such as lemmatization, eliminating common stopwords. The preprocessed data is then embedded to multiple different embedded vectors. We chose \textit{Word2Vec}, \cite{mikolov2013efficient} for focuses on neighboring words in a fixed-size and \textit{MiniLM BERT}, \cite{devlin2018bert} to  masked words based on the context. The advantage of the different embeddings assist us in keeping sentiment from BERT and word relation information from Word2Vec together for further enhanced training. These different embeddings are passed to the two identical arms of a siamese network consisting of Long Short Time Memory (LSTM), \cite{hochreiter1997long} learning algorithm. To qualify the predicted output, we pass it through fuzzy classifier, \cite{kuncheva2000fuzzy} for decision on the outcome. The figure below gives us a detail implementation of the network architecture.

\begin{figure}[h]
  \centering
  \includegraphics[width=\linewidth]{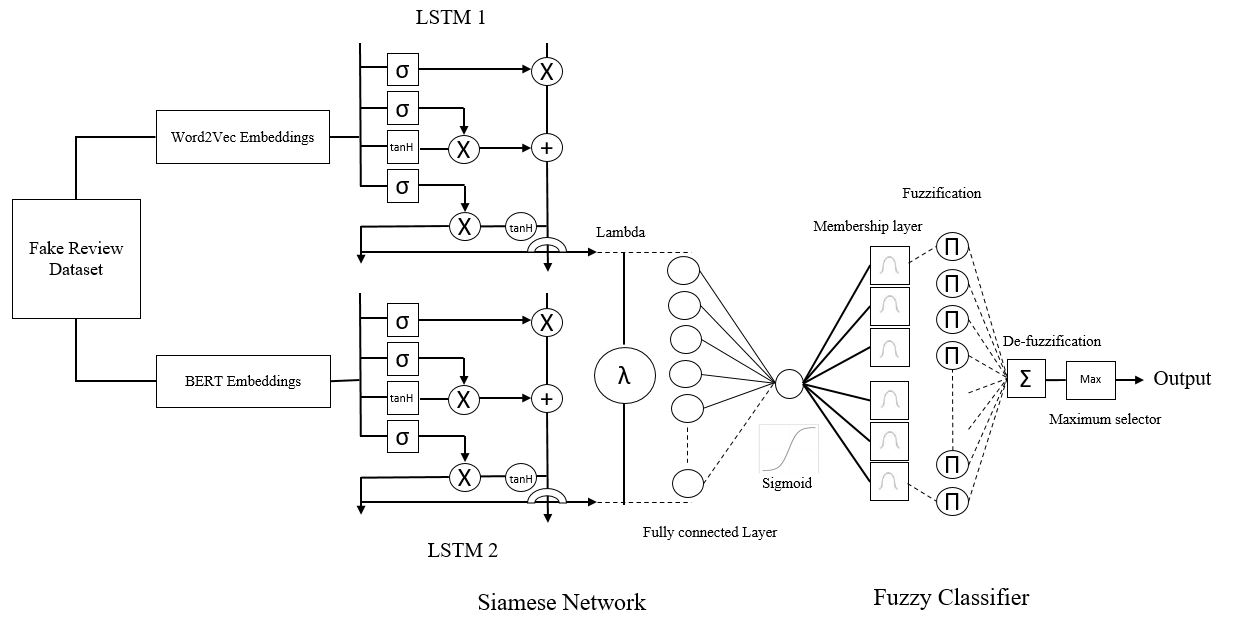}
  \caption{Our network architecture on fake review dataset}
   \label{fig:architecture}
\end{figure}

We have studied the complex nature of the reviews and considered Figure \ref{fig:architecture} as our approach for our algorithm. This approach might also be further developed for training networks in other natural language research area.

\section{Related Works}
Machine learning approach has contributed significantly in the field of natural language processing (NLP). \cite{torfi2021natural} gives us a detail discussion on the advancement on this field. Recurrent Neural Network (RNN) is one of the approach used for NLP based learning algorithm. And from \cite{yu2019review} we can derive that LSTM is one of the most common RNN at present. The LSTM network since its introduction has gone through multiple manipulation and enhancement as noted by \cite{LINDEMANN2021650}. We have also found that RNN such as LSTM specifically with its architecture is more suitable in a detection algorithm for fake reviews. Thus we put forward the RNN algorithm with LSTM for the prediction. As for the dataset, we surveyed many dataset and tried to build our own, however we found \cite{salminen2022creating} dataset with diverse categories and corpora an ideal candidate for stepping stone on the algorithm implementation. The user reviews are unclean with multiple punctuation and stopwords removal necessary process as per \cite{sun2014empirical} which might impact on dimension and feature extraction process. Previous works such as \cite{akhmetov2020highly} and \cite{plisson2004rule} discusses about importance of lemmatization in the diverse corpora to achieve required machine learning classifier. Stemming \cite{hull1996stemming} and stopwords removal \cite{silva2003importance} also assist in corpora cleaning to avoid loosing information on dimension reduction over vectorization.

Vectorization of the corpora is also a pre requirement process for input of machine learning in a siamese network \cite{chicco2021siamese}. The techniques such as Word2Vec, TF-IDF and Bag-of-Words \cite{krzeszewska2022systematic} are most common vectorization techniques. In Word2Vec \cite{al2019use} embeddings we found an ideal candidate, where the semantic relationship \cite{rong2014word2vec} are also taken into consideration. However, the Word2Vec fails to address the contextual relationship \cite{miaschi2020contextual} and hence loss of information is identified. BERT embeddings are more accurate in preserving the relationship in \cite{wiedemann2019does}. MiniLM BERT is a compact and efficient version for smaller corpora \cite{tsai2019small}. Thus, the other sub network could use the characteristics of MiniLM BERT embeddings \cite{wang2020minilm} of the identical dataset. These two sub network with the necessary information can be used for Siamese network. The network could be formed using a pairwise comparison as \cite{lv2020siamese} with LSTM layers as \cite{gleize2019you} for higher accuracy. On the neural network with text analysis could have imprecision \cite{aliev2006fuzzy} on the prediction. The decision boundaries on fuzzy classifier \cite{murmu2015application} add to the more robust decision making. This also gives a higher prediction accuracy \cite{ishibuchi1999performance} on uncertain data patterns such as reviews.

Thus, we can conclude that we could design a systems with a multi embedding convergence network which could compare and compete with state-of-the-art networks for classifications on the text reviews.

\section{Methodology}

The identification of fake reviews is a complex process. There is a limitation on diverse corpora and lack of efficient training algorithm. However, we have found  \cite{salminen2022creating} dataset, one of the meticulous datasets with diverse product reviews. The basic algorithm of the approach consist of eliminating the stopwords and punctuation which increases accuracy efficiency of the corpus. The next phase converts the word tokens to embeddings using MiniLM BERT transformer and Word2Vec. These embeddings acts as a sub network and passed through the training model of a cosine Siamese network. The output is then used as a input to the fuzzy logic classifier for most robust result on fake and real review detection. The approached algorithm steps can be subdivided into:
\begin{itemize}
\item {Understanding the dataset}  \item {Preprocess the corpora}  \item {Word embedding generation}  \item {Siamese training with LSTM}  \item {Fuzzy classifier decision output}
\end{itemize}

\subsection{Understanding the dataset}
The datasets consist of 40432 reviews of various products. The dataset includes columns of category, review, rating and label:"Fake" Computer generated (CG) or "Real" User given (OG). The details on each category are given below. The review is mostly on the e-commerce website product and the category on each product review is listed.

\begin{table}[ht]
  \caption{Yelp Dataset: Category-wise Fake and Real Reviews}
  \label{tab:reviews}
  \begin{tabular}{lcccc}
    \toprule
    Category & \multicolumn{2}{c}{Fake} & \multicolumn{2}{c}{Real} \\
    \cmidrule(lr){2-3} \cmidrule(lr){4-5}
    & No. of corpora & Average Length & No. of corpora & Average Length \\
    \midrule
    Books\_5 & 2185 & 374 & 2185 & 491 \\
    Clothing\_Shoes\_and\_Jewelry\_5 & 1924 & 250 & 1924 & 322 \\
    Electronics\_5 & 1994 & 306 & 1994 & 408 \\
    Home\_and\_Kitchen\_5 & 2028 & 272 & 2028 & 350 \\
    Kindle\_Store\_5 & 2365 & 375 & 2365 & 474 \\
    Movies\_and\_TV\_5 & 1794 & 335 & 1794 & 452 \\
    Pet\_Supplies\_5 & 2127 & 276 & 2127 & 354 \\
    Sports\_and\_Outdoors\_5 & 1973 & 282 & 1973 & 363 \\
    Tools\_and\_Home\_Improvement\_5 & 1929 & 292 & 1929 & 380 \\
    Toys\_and\_Games\_5 & 1897 & 276 & 1897 & 357 \\
    \midrule
    Total Reviews & 20216 & & 20216 & \\
    \bottomrule
  \end{tabular}
\end{table}

From Table \ref{tab:reviews}, we can observe that kindle store, books and pet supplies have more reviews than movies or toys bought. This might be because of the satisfaction of the buyer on the product or less accessibility on the product review page. Most of the time a satisfied or a product acceptance could lead to customer overlooking review options. One of the intriguing fact in the table is the average review length on each category. We can derive that fake reviews are smaller in length compared to the real reviews. This observation lead us to the opinion that fake reviews are more generic and not product detail oriented. There are multiple unwanted generic and emotional words which are also not of any assistance to decide on the fictitious nature of the reviews. Thus, we need to clean the corpus before any computational work is done on the dataset. We discuss that on the following preprocess section.

\subsection{Preprocess the corpora}
The data is passed through the standard corpora cleaning preprocess. This involves  punctuation removal, stopwords elimination and lemmatizing the tokens. The punctuation removal tool is available on standard nltk library. The accented characters in the corpora are also removed with unidecode library. Next phase is removing Common english stopwords, which can be done using nltk library. We found the most common words on the reviews includes article, conjunction and prepositions. These words are not considered informative in detecting fictitious reviews. Most detail oriented sentences as discussed in previous section, considering the corpus gives us an insight on the nature of the review. 

\begin{figure}[h]
  \centering
  \includegraphics[width=\linewidth]{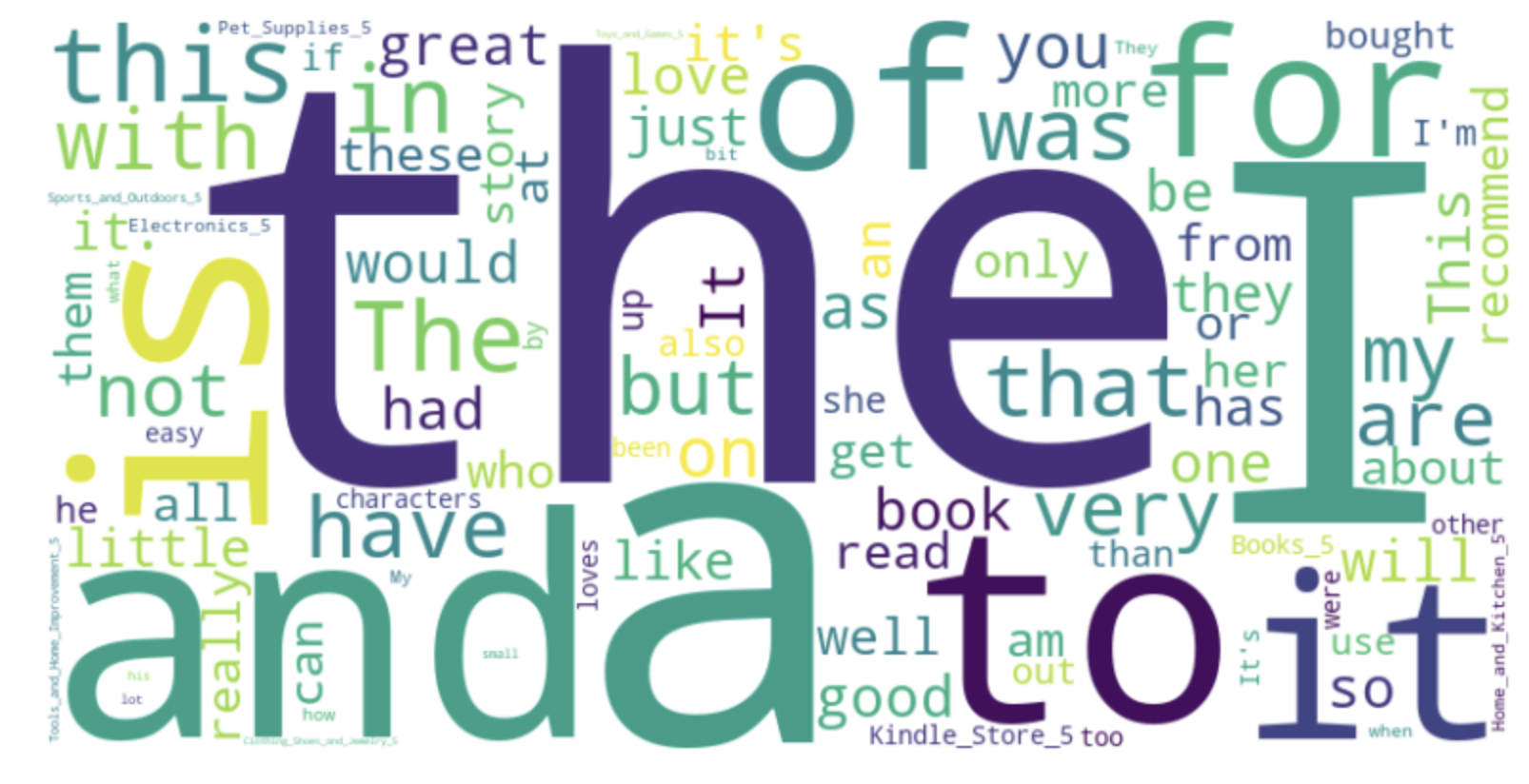}
  \caption{Word cloud for most common words found in Yelp Dataset}
  \label{fig:wordcloud}
\end{figure}

In Fig. \ref{fig:wordcloud}, we can observe the frequency of the most common words in the reviews. These words mostly cannot impact the review decision. The highest frequent word found in the data is 'the', followed by 'I', 'and', 'a' and furthermore. The frequency word cloud helps us understand the characteristics of sentences and assist us in eliminating words before vectors embeddings.

\subsection{Word embedding generation}

The preprocessed review data is then passed as tokens through two different embeddings. MiniLM BERT Transformer and Word2Vec.

\textit{Bidirectional Encoder Representations from Transformers (BERT)} is a transformer architecture for pre-training contextualized word representations. The architecture consist of encoder-decoder structure which allows to capture contextual relationship. The input stream of the MiniLM BERT encoder consist of word \textit{tokens}. The tokens are then converted into embeddings as a output with vector shape=(,384). The model can be used from the python library sentence-transformers

\textit{Word2Vec} is a vector representation, generating semantic relationship between review sentences. This helps us getting the features intact on the review corpus. The Word2Vec embeddings representation elevates the word tokens into higher dimensional space. The training module of Word2Vec consist of number of parameters below:
\begin{itemize}
\item {\verb|vector size|}: The dimensionality of the word vectors which is 384  
\item {\verb|window|}: The maximum distance between the current and predicted word within a corpus is 5   
\item {\verb|minimum count|}:  All words with a total frequency lower than 1  
\item {\verb|workers|}: The number of CPU cores to use when training is 5 
\end{itemize}

This will create an array vector of all the tokens in the dataset with the dimension of shape = (,384) 
\[  \mathbf{w_i} = \begin{bmatrix} upgrade \\ original \\ \vdots \\ year \end{bmatrix} \overrightarrow \mathbf{v(w_i)} =  \begin{bmatrix} v_1 \\ v_2 \\ \vdots \\ v_{384} \end{bmatrix} \]
where \(\mathbf{w_i}\) represents review on \(i\)-th row in the dataset, and \(\mathbf{v(w_i)}\) is the vector representation.

\begin{figure}[h]
  \centering
  \includegraphics[width=\linewidth]{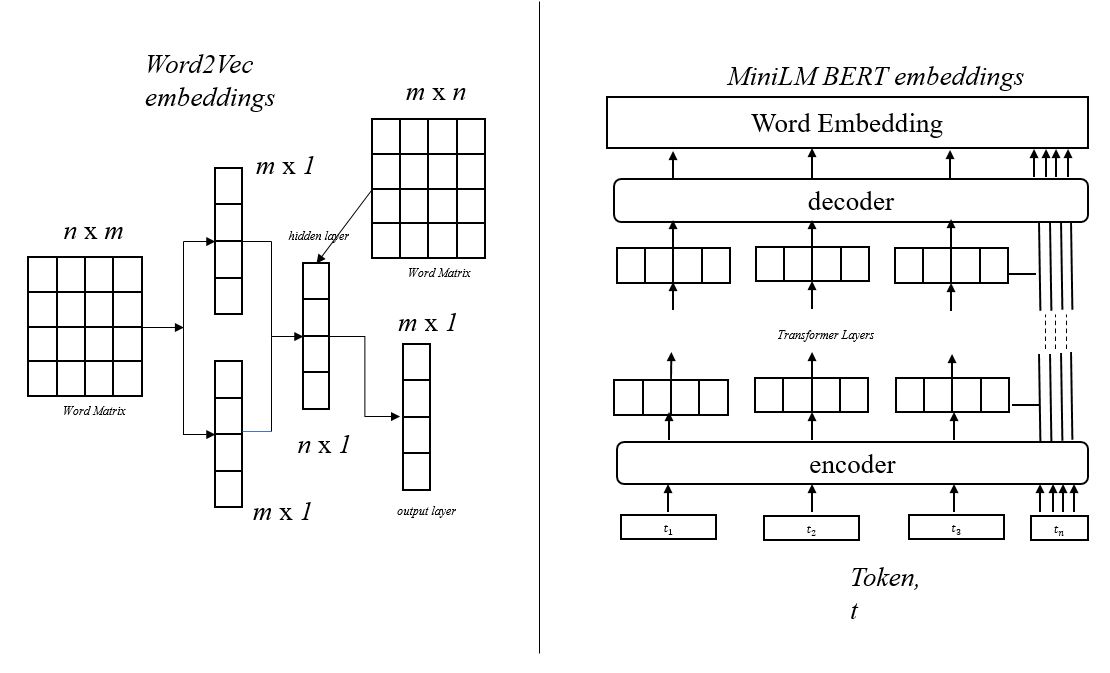}
  \caption{Word Embeddings: \textit{Left}: Word2Vec architecture, \textit{Right}: MiniLM BERT Transformer}
  \label{fig:Embeddings}
\end{figure}

\subsection{Siamese training with LSTM}
The Siamese network implemented in this architecture has two identical sub networks allowing to pass the MiniLM BERT and Word2Vec in parallel combination respectively. The layers consist of Long Short-Term Memory (LSTM) layer. This layer consist of three sub layers:
\textit{Input Gate ($i_t$):} Controls the update of the cell state with new information.
\textit{Forget Gate ($f_t$):} Controls the removal of information from the cell state.
\textit{Output Gate ($o_t$):} Controls the extraction of information from the cell state to produce the hidden state. The cell state in maintained on the input sequence from this layer.

The next few layers consist of matrix computation between both vector output of LSTM layer. This is followed by a cosine distance with \(\mathbf{L_2}\) normalization,
\[ \text{Cosine Distance}(\mathbf{A}, \mathbf{B}) = 1 - \frac{\mathbf{\hat{A}} \cdot \mathbf{\hat{B}}}{\|\mathbf{\hat{A}}\| \cdot \|\mathbf{\hat{B}}\|}
 \] where A and B are vectors.

The following layers consist of fully connected neural networks with dropout  and a binary classification layer with sigmoid activation function. The Siamese network is a two identical network implementation  \(f(\mathbf{x}_1)\) and \(f(\mathbf{x}_2)\) then passed through the cosine similarity metric. This network can be represented as, 
\[ \text{Similarity} = \frac{1 + \cos(f(\mathbf{x}_1), f(\mathbf{x}_2))}{2}
 \]
The model was compiled with \textit{binary crossentropy} and \textit{adam} optimizer with validation split of 0.3 and patience at 20. The final siamese output from the last layer is passed through the fuzzy logic classifier for decision making. 

\subsection{Fuzzy classifier decision output}
A fuzzy logic framework is used in decision-making procedures to enable the representation of rigorous decisions for imprecision and uncertainty. Based on the output of the siamese network features, fuzzy classifiers is used to assign membership functions degrees to the classes. To obtain membership values for the "Real" and "Fake" sets, membership functions are applied to the output. The link between the input features and the output classes is defined by a set of fuzzy rules. The input is fuzzified by applying fuzzy inference system procedures while accounting for the membership functions' assigned degrees of membership. Fuzzification creates fuzzy sets from input values. Every input value has a degree of membership attached to it. Aggregation is accomplished using the max function. Next, we determine the total membership. The process of defuzzification is used to transform the combined fuzzy output into a numerical value. A decision threshold is established, and whether or not the total membership exceeds, will determine the  decision.

\begin{figure}[h]
  \centering
  \includegraphics[width=\linewidth]{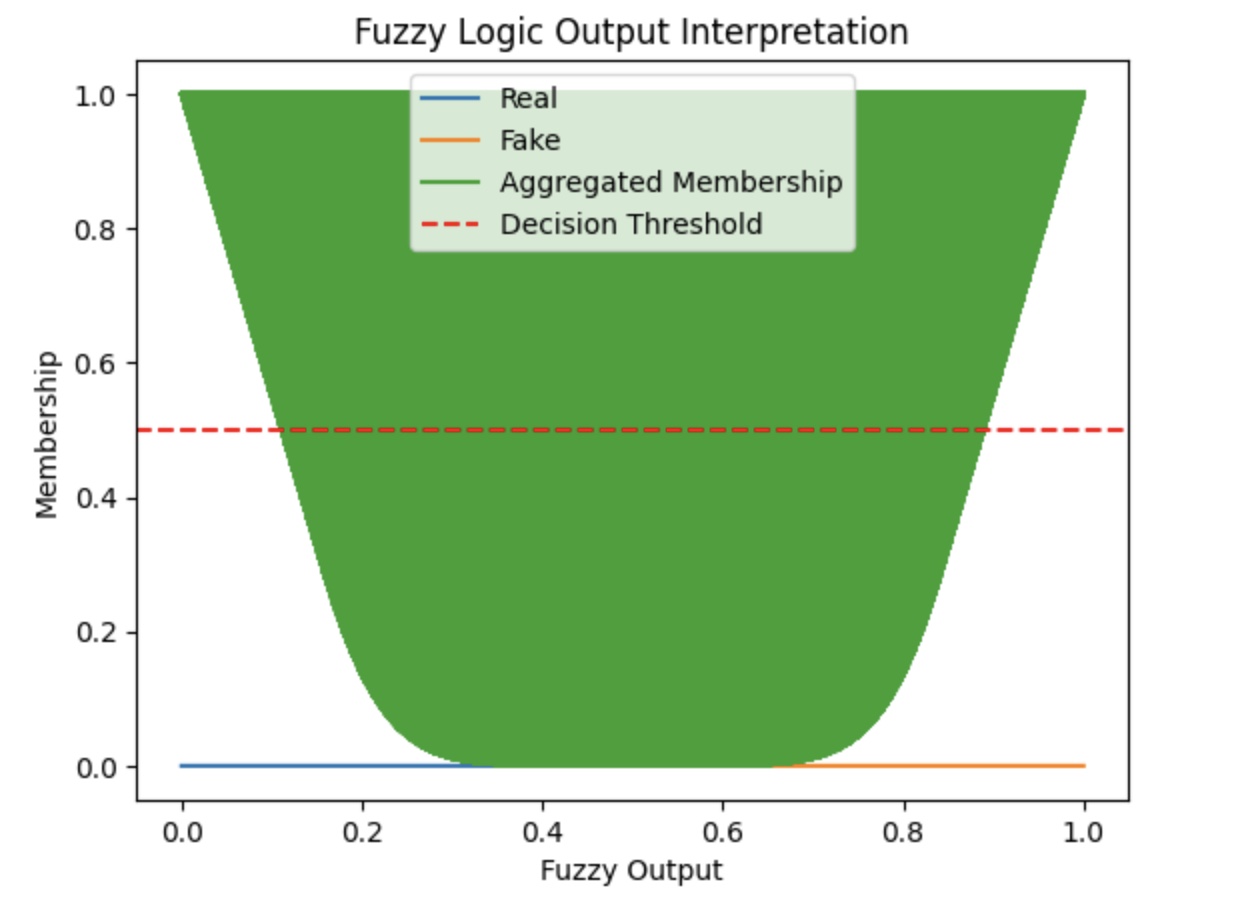}
  \caption{A Fuzzy Output for the Fake review detection on Siamese network}
  \label{fig:FuzzyOutput}
\end{figure}

\section{Results and Discussion}

In this section we discuss about our implementation results on the Siamese implementation on multi embeddings for our fake review dataset. The data is trained through the identical LSTM Siamese network for BERT and Word2Vec embeddings. These embeddings act as an input feature for identical siamese sub networks. Cosine distance is calculated and trained through a fully connected neural layers for binary classification. The results from the siamese training process has achieved an \textit{validation accuracy} of 84\%. The \textit{training accuracy} has reached 91\% on the mentioned dataset. These predicted output is sent through the fuzzy classifier with membership function. When passed through the fuzzy classifier, the prediction increases to 88\% on the decision making. These gives us a total of 35791 prediction from the model prediction on 40432 of the available corpuses. As shown in the figure \ref{fig:FuzzyOutput} we could identify the distribution of the review prediction on the Fuzzy classifier. Thus, we observe that there is a high boost on the fuzzy classifier on the predicted Siamese output.

\section{Future Works}
In the near future, we want to implement the robust networks in the YELP dataset \cite{yelp_dataset} to check the feasibility on the network efficiency and accuracy. The size of the YELP dataset is substantial. This may facilitate the extraction more information from the corpus with high vectorized dimension. The higher dimension might help us in more accurate results and better accuracy.Other embeddings that can be used as sub networks for a siamese architecture include TF-IDF, Bag of Words, and Word Mover's Distance. This may provide us with additional algorithmic solutions that are appropriate for various datasets and distinct corpora. These methods can be used in various domains like accessibility, security, and medical sciences, as well as computer vision and human-computer interaction. The new strategy of using a different embedding on the Siamaese module may lead to the development of more advanced networks. A finer adjustment to this architecture may provide new perspectives on the approaches presently in place for machine learning-based models.

\bibliographystyle{unsrtnat}
\bibliography{references}  






\end{document}